\documentclass[12pt]{iopart}
\usepackage{graphicx}
\usepackage{url}

\begin{document}
 
\title[Fast SPECT CDR Modeling]{Computationally Efficient Collimator-Detector Response Compensation in High Energy SPECT using 1D Convolutions and Rotations}

\author{Lucas A. Polson$^{1,2}$, Pedro Esquinas$^{3}$, Sara Kurkowska$^{2,4}$, Chenguang Li$^{1,2}$, Peyman Sheikhzadeh$^{5}$, Mehrshad Abbassi$^{5}$, Saeed Farzanehfar$^{5}$, Seyyede Mirabedian$^{5}$, Carlos Uribe$^{2,3,6}$, Arman Rahmim$^{1,2,6}$}

\address{$^1$ Department of Physics \& Astronomy, University of British Columbia, Vancouver, Canada}
\address{$^2$ Department of Integrative Oncology, BC Cancer Research Institute, Vancouver Canada}
\address{$^3$ Molecular Imaging and Therapy Department, BC Cancer, Vancouver Canada}
\address{$^4$ Department of Nuclear Medicine, Pomeranian Medical University, Szczecin, Poland}
\address{$^5$ Nuclear Medicine Department, IKHC, Faculty of Medicine, Tehran University of Medical Science, Tehran, Iran}
\address{$^6$ Department of Radiology, University of British Columbia,Vancouver, Canada}

\ead{lukepolson@outlook.com}

\begin{abstract}
\textit{Objective}: Modeling of the collimator-detector response (CDR) in SPECT reconstruction enables improved resolution and accuracy, and is thus important for quantitative imaging applications such as dosimetry. The implementation of CDR modeling, however, can become a computational bottleneck when there are substantial components of septal penetration and scatter in the acquired data, since a direct convolution-based approach requires large 2D kernels. This work proposes a 1D convolution and rotation-based CDR model that reduces reconstruction times but maintains consistency with models that employ 2D convolutions. To enable open-source development and use of these models in image reconstruction, we release a SPECTPSFToolbox repository for the PyTomography project on GitHub. \textit{Approach}: A 1D/rotation-based CDR model was formulated and subsequently fit to Monte Carlo point source data representative of ${}^{177}$Lu, ${}^{131}$I, and ${}^{225}$Ac imaging. Computation times of (i) the proposed 1D/rotation-based model and (ii) a traditional model that uses 2D convolutions were compared for typical SPECT matrix sizes. Both CDR models were then used in the reconstruction of Monte Carlo, physical phantom, and patient data; the models were compared by quantifying total counts in hot regions of interest (ROIs) and activity contrast between hot ROIs and background regions. \textit{Results}: For typical matrix sizes in SPECT reconstruction, application of the 1D/rotation-based model provides a two-fold computational speed-up over the 2D model when running on GPU. Only small differences between the 1D/rotation-based and 2D models (order of $1\%$) were obtained for count and contrast quantification in select ROIs. \textit{Significance}: A technique for CDR modeling in SPECT was proposed that (i) significantly speeds up reconstruction times, and (ii) yields nearly identical reconstructions to traditional 2D convolution based CDR techniques. The released toolbox  will permit open-source development of similar models for different isotopes and collimators.
\end{abstract}

%
%
%
%
%

\section{Introduction}

Single photon emission computed tomography (SPECT) is an \textit{in vivo} imaging modality, possessing value for various clinical applications \cite{cardiac_spect, brain_spect, bone_spect}. In particular, quantitative SPECT imaging has applications in the field of theranostics, wherein image quantification and dosimetry can be performed to improve radiopharmaceutical therapies (RPTs). Routine SPECT-based dosimetry could unveil the relationships between tumor-response and healthy organ complications with absorbed dose, therefore enabling personalized treatments that maximize tumor dose while minimizing toxicity to organs at risk \cite{RPT, Kiess2024}. As a recent and well-known example, the success of ${}^{177}$Lu-PSMA-617 based RPTs in the VISION \cite{lu177_trial1} and TheraP \cite{lu177trial2} randomized control trials has lead to expanded research and clinical translation efforts for wide-scale deployment, as well as pursuit of absorbed dose estimation as a tool towards personalized treatments \cite{lu1,lu2}. 

SPECT-based dosimetry relies on accurate quantification of activity distribution from SPECT images, which is strongly dependent on the system model used in image reconstruction. The system model is a linear operator that predicts the expectation of the acquired data given a 3D isotope distribution, accounting for phenomena such as photon attenuation in the patient. An important aspect of the model is collimator detector response (CDR) modeling, which estimates image blurring caused by the collimator and the detector \cite{Frey2006, metz1980geometric, septal}. Computation of the CDR is typically the most computationally expensive operation in image reconstruction.

The CDR can be characterized by the system's response to a point source of activity; this is denoted the point spread function (PSF). The PSF can be decomposed into multiple components. The intrinsic response function (IRF) characterizes the uncertainty on the point of interaction within the detecting material (e.g. an NaI scintillator), and is sufficiently modeled using a Gaussian function. The collimator response for parallel hole collimators results from the inability of the collimator to accept only photons travelling perpendicular to the detector; it consists of three components: (i) the geometric response function (GRF) \cite{metz1980geometric, GRF_2, GRF_3, GRF_4}, which describes photons that travel through the collimator holes without penetrating or interacting with the septa, (ii) the septal penetration response function (SPRF), which describes the contribution from photons that travel through the collimator without being attenuated, and (iii) the septal scatter response function (SSRF), which consists of photons that interacted and scattered within the collimator and were subsequently detected. 

The selection of collimator parameters is an important aspect of SPECT imaging \cite{spect_collimator_design}. A collimator with longer, narrower bores, and thicker septal thickness reduces the contribution from the SPRF and SSRF by increasing the probability of attenuation for photons not traveling perpendicular to the detector. If the collimator geometry is optimized as such, the net point spread function (PSF) is dominated by the IRF + GRF and can be reasonably approximated using a 2D Gaussian function.  A trade-off of having a thicker collimator with smaller hole diameters, however, is a decrease in detector sensitivity; the corresponding implications for quantitative imaging is a decrease in precision or longer patient scan times. This trade-off must be independently considered for each isotope. For ${}^{177}$Lu labeled radiopharmaceuticals, where the $208~$keV photopeak is typically used for imaging, a commercially labeled ``medium energy'' collimator configuration (i) yields a reasonable count rate and (ii) adequately minimizes the SPRF and SSRF components. In this situation, a 2D Gaussian PSF model is sufficient for image reconstruction \cite{Uribe2017, Ljungberg151}.

To the knowledge of the authors, all presently available commercial reconstruction software offer Gaussian PSF modeling, and thus implicitly assume the SPRF and SSRF are negligible. There are multiple advantages to this. Firstly, the computational advantage of 2D Gaussian PSF modeling is that 2D Gaussian convolution is separable into use of two perpendicular 1D Gaussian convolutions, which are less computationally expensive to implement. Secondly, the Gaussian PSF model used to model the GRF+IRF can be obtained analytically for any photon emission energy and standard collimator shapes, and thus does not require lookup tables. However, when the PSF has significant contributions from the SPRF and SSRF, 2D Gaussian PSF modeling fails to capture all the features of the PSF. This is typically an issue with radioisotopes that emit high photon energies, such as $\alpha$-emitters like ${}^{225}$Ac.

Use of $\alpha$-emitters in radiopharmaceutical therapies presents an exciting frontier due to the high linear energy transfer (LET) associated with $\alpha$ particles, which may yield benefits for tumor eradication \cite{ac1,ac2,ac3,ac4,ac5,ac6}. In preclinical studies, simultaneous treatment with ${}^{177}$Lu-PSMA-617 and ${}^{225}$Ac-PSMA-617 compared to ${}^{177}$Lu-PSMA-617 alone resulted in significantly reduced tumor growth \cite{Meyerjnumed.123.265433}. Kratochwil et al.\ \cite{kratochwil2016ac} applied ${}^{225}$Ac-PSMA-617 to patients with metastatic castration resistant prostate cancer (mCRPC) who previously exhausted ${}^{177}$Lu-PSMA-617 treatment. Certain patients achieved a full response, with prostate specific antigen (PSA) levels in one patient decreasing from 419 ng/mL to below 0.1 ng/mL. At the time of writing, there are ongoing studies of ${}^{225}$Ac based radiopharmaceuticals looking at dose escalation \cite{dose_esc1,dose_esc2}, fractionation \cite{dose_frac}, and safety and efficacy \cite{safety_ac225}. A recent meta analysis \cite{ac225_meta} elaborates further on these clinical trials, discussing toxicities and other challenges with these treatments. Quite recently, targeted $\alpha$ therapy with ${}^{213}$Bi has been shown to reduce amyloid plaque concentration in male mice \cite{bi213_amyloid}, eluding to a potential treatment option for Alzheimer disease.

Throughout the decay chain of ${}^{225}$Ac, the daughters ${}^{213}$Bi and ${}^{221}$Fr emit photons detectable within a SPECT system of $440~$keV and $218~$keV respectively. Unfortunately, even with the  commercially available collimators designed for high-energy photons, there are still significant SPRF and SSRF components present in the ${}^{213}$Bi $440~$keV peak; a sample PSF is shown in Figure \ref{fig:psf_description}. This phenomena similarly occurs in imaging of photons with energies of 511~keV (positron emitter) \cite{psf_511} and $364$~keV (${}^{131}$I) \cite{Chun2013}. As a consequence, the PSF can no longer be modeled using a 2D Gaussian and is thus no longer separable into two perpendicular 1D components. Since this is the bottleneck of system modeling, image reconstruction takes significantly longer. 

The reduction of reconstruction times in medical imaging remains an important research topic. Tsai et al.\ \cite{Tsai2018} recently showed that the limited-memory Broyden-Feltcher-Goldfarb-Shannon algorithm with box constraints and a diagonal preconditioner (L-BFGS-B-PC) was able to converge several times faster than the one step late expectation maximum (OSL-EM) algorithm in Positron Emission Tomography (PET) reconstruction because less projection operations were required. As another example, Chun et al.\ \cite{Chun2013} reduced reconstruction times of ${}^{131}$I by using fast fourier transform (FFT) based 2D convolutions for PSF modeling. By contrast, in the present work, we seek to reduce reconstruction times by using a PSF model that incorporates (i) 1D convolutions and (ii) rotations. It will be shown that this model runs faster than standard and FFT based 2D convolutions when implemented on a GPU, and can thus be used to speed up image reconstruction times. The method is evaluated on Monte Carlo phantom data of ${}^{177}$Lu, ${}^{131}$I, and ${}^{225}$Ac, as well as physical phantom and patient data of ${}^{225}$Ac. In each case, the method demonstrates near-identical recovery coefficients and contrast in regions of interest compared to MC-based 2D convolution techniques.

Alongside this paper, we release the SPECTPSFToolbox: an open-source GitHub repository which forms a new component of the PyTomography \cite{pytomo} project. The toolbox contains functionality for developing and fitting arbitrary PSF models to arbitrary point-source data. The saved models can then be loaded in our in-house initiated and community developed library PyTomography for SPECT reconstruction. As such, the techniques used in this paper can also be applied to other isotope / collimator configurations that are of interest in the nuclear medicine community. To encourage community use, we have released nine tutorials demonstrating how to use the toolbox, and how to integrate the models in PyTomography. The link to the PyTomography project (which includes the SPECTPSFToolbox) is \url{https://github.com/PyTomography}

\section{Materials and Methods}\label{sec2}

Section 2.1 outlines the mathematical formalism of the PSF model and establishes the reconstruction protocol for all acquired data presented in subsequent sections. Section 2.2 describes how the model is fit to Monte Carlo (MC) ${}^{177}$Lu, ${}^{131}$I, and ${}^{225}$Ac point source data acquisitions simulated at various source-detector distances.  The computational time of the model is benchmarked and compared to 2D PSF modeling and Gaussian PSF modeling. In Section 2.3, the model is then used for reconstruction of (i) MC simulated ${}^{177}$Lu, ${}^{131}$I, and ${}^{225}$Ac phantoms, (ii) a physical ${}^{225}$Ac phantom, and (iii) a patient receiving ${}^{225}$Ac-PSMA-617 treatment for metastatic prostate cancer. All computation was performed using a Microsoft Azure virtual machine (Standard NC6s v3) with a 6 CPUs (Intel(R) Xeon(R) CPU E5-2690 v4 @ 2.60GHz), 112 GB of RAM, and a Tesla V100 GPU.

All MC data was generated using SIMIND \cite{simind}. In order to capture the effects of septal penetration and scatter, SIMIND has an optional routine based on the delta-scattering technique \cite{delta_scatter}, which has been validated for high-energy isotopes such as ${}^{131}$I \cite{simind_validation_i131}. This SIMIND routine has been used in previous studies to generate stacks of PSFs at various source-detector distances for MC-based ${}^{225}$Ac PSF modeling in custom reconstruction \cite{ac5}, since the reconstruction software of the scanner manufacturers does not provide comprehensive PSF modeling of high energy isotopes. 

\subsection{Theory}

\subsubsection{PSF Modeling}

In this work, the notation $f(x,y;d)$ is used for 3D objects: $d$ denotes the distance between a plane parallel to the detector and the detector, and $(x,y)$ denote the position on the plane. It is assumed that $x$, $y$, and $d$ are discrete and thus $f$ consists of voxels. The following notation is used to represent convolution operator $K$:

\begin{eqnarray}
    K^{(x;d;b)}f \equiv \sum_{x'} f(x',y;d) k(x-x';d;b)\\
    K^{(y;d;b)}f  \equiv \sum_{y'} f(x,y';d) k(y-y';d;b)\\
    K^{(x,y;d;b)}f \equiv \sum_{x',y'} f(x',y';d;b) k(x-x',y-y';d;b) \label{eq:op_form_2D}
\end{eqnarray}
where $b$ are additional parameters the kernel $k$ depends on, such as collimator septal thickness $L_b$, hole diameter $w_b$, and the linear attenuation coefficient of the collimator material $\mu_b$.

Assuming a linear shift invariant (LSI) PSF, the SPECT system matrix estimates the projection $g_{\phi}$ at angle $\phi$ as

\begin{equation}
    g_{\phi}(x,y) = \sum_{d} \sum_{x',y'} k_{\mathrm{PSF}}(x-x',y-y',d) p_{\mathrm{att}}(x,y,x',y',d) f(x,y,d)
    \label{eq:model_full}
\end{equation}
where $(x',y')$ is the source position on a plane parallel to the detector in 3D space, $\mathrm{psf}(...,d)$ is a kernel that yields the point spread function (PSF) at a distance $d$ from the detector, and $p_{\mathrm{att}}(x,y,x',y',d)$ is the probability that photons traveling from $(x,y,d)$ to detector coordinate $(x', y')$ are not attenuated. If the attenuation probabilities are approximately constant for different photon impingement angles permitted on the detector (e.g.\ valid when the PSF is small), it follows that $p_{\mathrm{att}}(x,y,x',y',d)$ can be replaced with $p_{\mathrm{att}}(x,y,x,y,d)$, which can simply be written as $p_{\mathrm{att}}(x,y,d)$. Under this assumption, $f'\equiv p_{\mathrm{att}}f$ can be defined as the ``attenuation-adjusted'' image, and Equation \ref{eq:model_full} can then be rewritten in operator form (Equation \ref{eq:op_form_2D}) as

\begin{equation}
    g_{\phi}(x,y) = \sum_{d} K_{\mathrm{PSF}}^{(x,y;d;b)} f'
    \label{eq:model_approx}
\end{equation}
Since this convolution operation often forms the bottleneck of SPECT system matrix modeling and image reconstruction, it is of interest to look for techniques to reduce the computation time. Under conditions of no septal penetration and scatter, the CDR is dominated by the GRF: $ K_{\mathrm{PSF}}^{(x,y;d;b)}$ can then be sufficiently approximated using a 2D Gaussian convolution $K_{\mathrm{2DG}}^{(x,y;d;b)}$ where the kernel is given by

\begin{equation}
    k_G(x,y;d;L_b,w_b,\mu_b) = \frac{1}{2 \pi \sigma(d,L_b,w_b,\mu_b)^2} \mathrm{exp} \left(-\frac{x^2+y^2}{2\sigma(d,L_b,w_b,\mu_b)^2} \right)
    \label{eq:gaussian_psf}
\end{equation}
\begin{equation}
    \sigma(d,L_b,w_b,\mu_b) =  \frac{1}{2\sqrt{2\log(2)}} \left(\frac{w_b}{L_b-2/\mu_b} \cdot d + w_b \right)
\end{equation}

\begin{figure}
\centering
\includegraphics[width=1.0\textwidth]{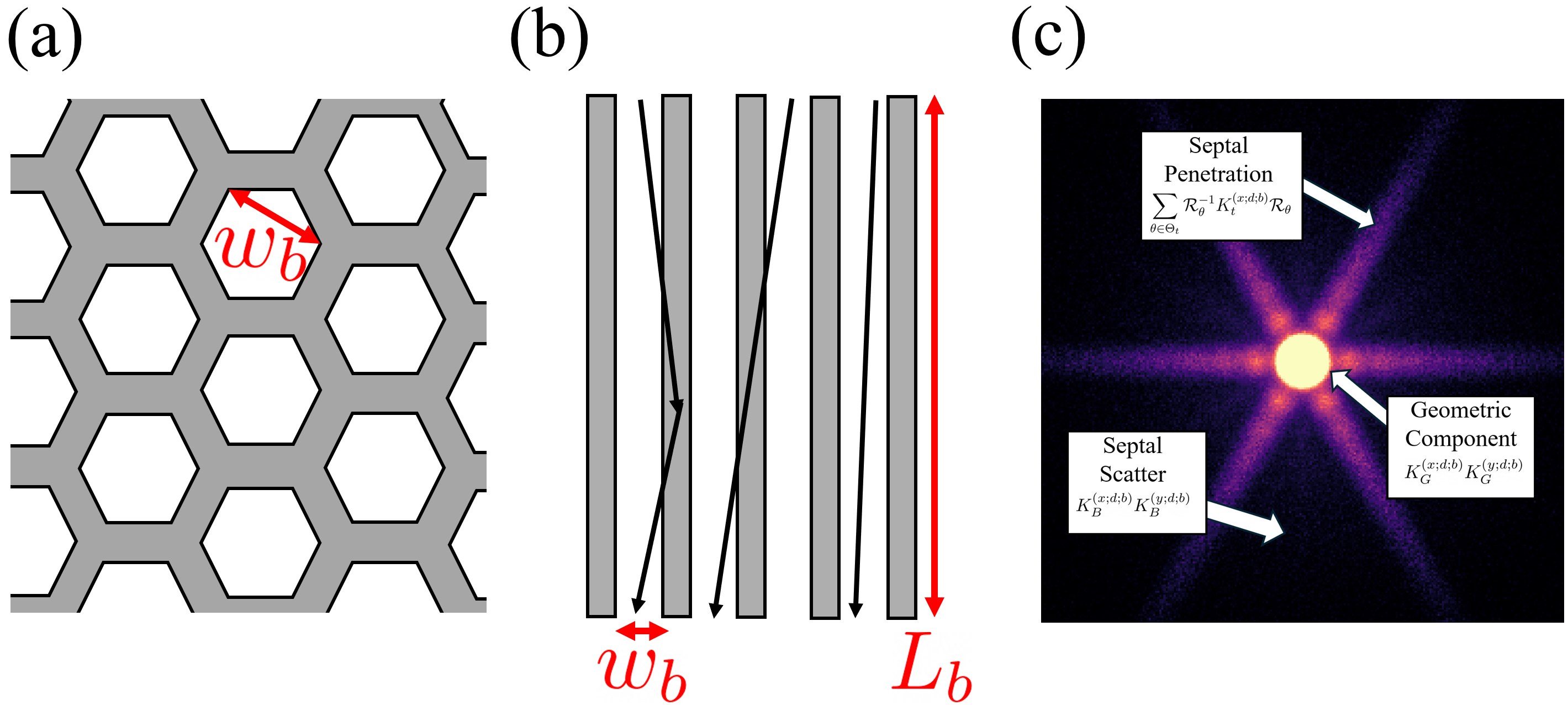}
\caption{(a) Top view of collimator, where $w_b$ is the collimator hole width. (b) Side view of collimator, where $L_b$ is the length of the collimator. Three different photon detection paths are shown as black arrows; from left to right they correspond to septal scatter, septal penetration, and geometric component respectively. (c) 440~keV PSF obtained by simulating a point source at a source-detector distance of 30~cm using SIMIND. Labeled scatter components that refer to the different terms of Equation \ref{eq:1dr} are shown.}\label{fig:psf_description}
\end{figure}

Convolution with a Gaussian function has computational advantages since it can be decomposed into successive application of two perpendicular 1D kernels via $K_{\mathrm{2DG}}^{(x,y;d;b)}=K_{\mathrm{1DG}}^{(x;d;b)}K_{\mathrm{1DG}}^{(y;d;b)}$; 1D convolution is significantly more computationally efficient than 2D. Unfortunately, the decomposition of a 2D kernel into two 1D kernels is not mathematically possible when the CDR contains significant SPRF and SSRF components. 2D convolution, however, is not the only way to implement Equation \ref{eq:model_approx}. Owing to the discrete rotational symmetries and features of the anisotropic PSFs obtained with parallel hole SPECT collimators, we propose the following ``1D rotation'' model (abbreviated as ``1DR'')  for the PSF operator:

\begin{equation}
    M_{\mathrm{1DR}}^{(x,y;d;b)}  \equiv  \left(\sum_{\theta \in \Theta_t}  \mathcal{R}^{-1}_{\theta} K^{(x;d;b)}_t \mathcal{R}_{\theta} + K_B^{(x;d;b)} K_B^{(y;d;b)} + 1\right) K^{(x;d;b)}_G K^{(y;d;b)}_G \label{eq:1dr}
\end{equation}
where $\mathcal{R}_{\theta}$ is a rotation operator that implements rotation about an axis in the $d$ direction by angle $\theta$, and $\Theta_t$ correspond to the angle of the septal penetration tails (equal to $\{0,\pi/3,2\pi/3\}$ for hexagonal collimators). This derived linear operator only makes use of 1D convolutions and rotations, since they require less computational time on GPU compared to 2D convolutions. The kernels, modeling the components of the CDR (as shown in Figure \ref{fig:psf_description}), are selected as follows:

\begin{enumerate}
    \item Gaussian kernel $ K^{(x;d;b)}_G$:
        \begin{equation}
            k_G(x;d;b) = A_G(d,b) \cdot \mathrm{exp} \left(-x^2 / 2\sigma_G(d,b)^2 \right)
            \label{eq:gauss_comp}
        \end{equation}
        \begin{equation}
            A_G(d,b) = b_0e^{-b_1d} + b_2e^{-b_3 d}
            \label{eq:amplitude}
        \end{equation}
        \begin{equation}
            \sigma_G(d,b) = b_4 + b_5\left(\sqrt{d^2+b_6^2} - |b_6|\right)
        \end{equation}
    This kernel is used to build geometric component of the PSF. The terms $A_G(d,b)$ and $\sigma_G(d,b)$ are able to capture the decaying amplitude and spreading out of the PSF at larger distances $d$.
        
    \item Tail kernel $ K^{(x;d;b)}_t$:
        \begin{equation}
            k_t(x;d;b) = A_t(d,b) f_t(x/\sigma_t(d,b))
            \label{eq:tail_comp}
        \end{equation}
        \begin{equation}
            A_t(d,b) = b_7e^{-b_8d} + b_9e^{-b_{10} d}
        \end{equation}
        \begin{equation}
            \sigma_t(d,b) =1+b_{11}\left(\sqrt{(d-d_{\mathrm{min}})^2+b_{12}^2} - |b_{12}| \right)
        \end{equation}
    where $d_{\mathrm{min}}$ is the source-detector distance used in the PSF fit. $f_t(x)$ is a discrete array of numbers that is linearly interpolated between its fixed points. This kernel models the tails of the septal penetration PSF component. Similar to the geometric component, the terms $A_t(d,b)$ and $\sigma_t(d,b)$ are such that the amplitude decays and the component spreads out at larger distances $d$. The function $f_t(x/\sigma_t(d,b))$ is initialized as a mono-exponential function to match the decay of the tails observed in Figure \ref{fig:psf_comp}.
    
    \item Isotropic background kernel $ K^{(x;d;b)}_B$:
        \begin{equation}
            k_B(x;d;b) = A_B(d,b) f_B(x/\sigma_B(d,b))
            \label{eq:bkg_comp}
        \end{equation}
        \begin{equation}
            A_B(d,b) = b_{13}e^{-b_{14}d} + b_{15}e^{-b_{16} d}
        \end{equation}
        \begin{equation}
            \sigma_B(d,b) =1+b_{17}\left(\sqrt{(d-d_{\mathrm{min}})^2+b_{18}^2} - |b_{18}| \right)
        \end{equation}
        \begin{equation}
            f_B(x) = e^{-|x|}
        \end{equation}
        This kernel is used to build the septal scatter component of the PSF. Similar to the previous two components, the terms $A_B(d,b)$ and $\sigma_B(d,b)$ capture the amplitude decay and spreading out of the component at larger distances $d$. The function form of $f_B(x)$ was selected based on the approximate decay of septal scatter seen in Figure \ref{fig:psf_comp}.
\end{enumerate}

All three model components also encapsulate the small blurring contribution from the IRF. The hyperparameters $b$ are separate for each of the three different components.

PSF modeling can also be accomplished using a generic 2D kernel; the model can be expressed as

\begin{equation}
    M_{\mathrm{2D}}^{(x,y;d;b)} \equiv K_{\mathrm{2D}}^{(x,y;d)} \label{eq:2d}
\end{equation}
where the corresponding kernel $k_{\mathrm{2D}}^{(x,y;d)}$ might be obtained from real data measurements, MC measurements, or an analytical model based on measurements (such as in \cite{Tsai2018}). In subsequent sections, the following three methods are compared:

\begin{enumerate}
    \item The 2D Gaussian model of Equation \ref{eq:gaussian_psf}, with parameters analytically obtained from collimator dimensions. Denoted ``2DG'', this is the technique used by available commercial reconstruction software. 
    \item A generic 2D Monte Carlo model (Equation \ref{eq:2d}) obtained directly from a stack of normalized point source MC data at various source-detector distances. Denoted ``2D-MC'', this is the same technique used by Delker et.\ al.\ \cite{ac5} for reconstruction of ${}^{225}$Ac patient data; it will be used as a standard against which method (iii) is compared.
    \item The proposed 1DR model (Equation \ref{eq:1dr}) with parameters obtained via fitting to MC point source data, and denoted ``1DR-MC''. Since it is fit to the data used for method (ii), an ideal model would yield identical reconstructions to those obtained with method (ii).
\end{enumerate}

It should be emphasized that method (i) does not require any empirical data, while methods (ii) and (iii) use MC point source data (hence the inclusion of ``-MC'' in the acronyms). Methods (ii) and (iii), however, can also be implemented using real point source measurements if that data is available.

The publicly available SPECTPSFToolbox repository is structured to facilitate the customizability of these models. For the 1DR-MC model, components are implemented via separate class instances that are subsequently added and multiplied together to obtain the final form of Equation \ref{eq:1dr}. The functional form and hyperparameters of the amplitude/scaling, (e.g.\ Equation \ref{eq:amplitude}) are left arbitrary for the user; this permits full customization of PSF models. The 2D-MC model is implemented via a class instance that receives a stack of PSF kernels with associated source-detector distances: when used in image reconstruction, the model automatically chooses the PSF kernel closest to each source-detector distance in the reconstruction problem.

\subsubsection{Image Reconstruction}

All acquired image data in subsequent sections are reconstructed using PyTomography with the ordered subset expectation maximization algorithm (OSEM); when one subset is used, the algorithm is denoted as maximum likelihood expectation maximization (MLEM). System matrices employed attenuation correction using attenuation maps derived from either (i) ground truth attenuation maps from MC simulations or (ii) acquired computed tomography (CT) images in physical studies. The imaging system equation is given by

\begin{equation}
    \bar{y} = Hx + \hat{s}
    \label{eq:image_system}
\end{equation}
where $\bar{y}$ is the expectation of the acquired count data $y$, $H$ is the system matrix, and $x$ is the estimated 3D count distribution. $\hat{s}$ is an additive term that accounts for phantom or patient scatter, events that back-scattered in the detector from higher energy emission peaks, and stray background radiation; the presence of stray background radiation for low count SPECT has been described by Zekun et al.\ \cite{abi}. For all cases in this paper, the triple energy window (TEW) technique is used to estimate $\hat{s}$.

When used in image reconstruction, the size of the 2D PSF kernels in the 2D-MC method was always set to $N-1 \times N-1$ where $N$ is the number of voxels along the largest direction in the reconstructed image. For the 1DR-MC method, the kernels corresponding to $K_G$ and $K_B$ were of size $N-1$, while the kernel corresponding to $K_t$ was of size $\lceil \sqrt{2} N \rceil + n$ to account for the diagonal, where $n$ is of integer value 0 or 1 to make the kernel size odd. It should be noted that for PSFs much larger than the dimensions of the image, the size of the kernels would need to cover a $2N-1 \times 2N-1$ area to account for contributions from voxels on one corner of the image to detector elements on the opposite corner; in practice, the PSFs were not that large and the dimensions chosen were sufficient.

\subsubsection{Image Quality Metrics}

To compare images reconstructed using the three different PSF models, two image metrics are used. The first metric is recovery coefficients (RCs), defined as
\begin{equation}
    \mathrm{RC} \equiv \frac{A_{\mathrm{pred}}}{A_{\mathrm{true}}}
\end{equation}
where $A_{\mathrm{pred}}$ corresponds to the predicted activity concentration in a region of interest (ROI), and $A_{\mathrm{true}}$ corresponds to the true activity in that region. In MC simulations consisting of spheres with infinite resolution, the ROIs on the reconstructed voxel grid are segmented such that voxels inside ROI are assigned 1 while voxels outside are assigned 0. The calibration factors necessary to convert counts to activity units are obtained via point source simulations with the same detector parameters. For the physical data, where calibration factors are not available, average counts are instead reported. The second metric used is contrast, defined as 

\begin{equation}
    \mathrm{Contrast} \equiv \frac{C_{\mathrm{hot}}}{C_{\mathrm{bkg}}}
\end{equation}
where $C_{\mathrm{hot}}$ is the number of counts in a hot ROI and $C_{\mathrm{bkg}}$ is the number of counts in the background. 

\subsection{Validation and Timing}

Point sources with energies of 208~keV, 364~keV, and 440~keV (representative of the typical primary energies acquired in ${}^{177}$Lu, ${}^{131}$I, and ${}^{225}$Ac imaging respectively) were simulated using the SIMIND Monte Carlo program \cite{simind} at 1100 positions that linearly varied from 0~cm to 58.44~cm. The detector pixel size was $0.24~\mathrm{cm} \times 0.24~\mathrm{cm}$ with $255\times255$ pixels. For ${}^{177}$Lu, the simulated collimator corresponded to a Siemens low energy high resolution (LEHR) configuration. While typical clinical protocols use a medium energy (ME) configuration to minimize septal penetration and scatter, the choice of collimator here is merely to demonstrate the feasibility of using a LEHR collimator with an advanced PSF model in reconstruction. A Siemens high energy (HE) configuration was used for ${}^{131}$I and ${}^{225}$Ac, which is standard for clinical protocol. The intrinsic resolution of the detector was also included in the simulation and was assumed to be $0.38~$cm at 140~keV, representative of a Symbia system with $3/8$'' crystal length. The collimator cover and backscatter components of the detector were modeled via 0.1~cm of aluminum and 6.6~cm of pyrex respectively; inclusion of these parameters is necessary to capture further scattering of detected photons present in real systems.

Twelve of the simulated PSFs per isotope/collimator configuration (at source-detector distances of $1~$cm and every $5~$cm from $5~$cm to $55~$cm) were used as data to fit the 1DR-MC model. Fitting was performed in three steps: (i) the Gaussian parameters were optimized using the Adaptive Moment Estimation (ADAM) algorithm for $1 \cdot 10^4$ iterations and a learning rate of $10^{-2}$, (ii) all the parameters were simultaneously optimized using ADAM for $1.5 \cdot 10^4$ iterations and a learning rate of $10^{-3}$. The PSFs at all distances were used to build the 2D-MC models.

The computational time for the 1DR-MC and 2D-MC models were bench-marked on CPU and GPU for by applying the ${}^{225}$Ac operators to matrices of four different sizes: $64^3$, $128^3$, $196^3$, and $256^3$. Each matrix of size $N^3$ was filled with random uniform numbers between 0 and 1. Each experiment used conditions similar to SPECT imaging, where the $d$ axis varied from $0~$cm to $50~$cm, and the $x$ and $y$ axes varied from $-30.74~$cm to $30.74~$cm. Use of standard convolution and fast fourier transform (FFT) based convolution were compared for each method. 

\subsection{SPECT Studies}

\subsubsection{Monte Carlo Phantom Studies}

To demonstrate the versatility of the proposed method, application of the 2DG, 2D-MC, and 1DR-MC models were evaluated for reconstruction of MC simulated ${}^{177}$Lu, ${}^{131}$I, and ${}^{225}$Ac projection data. Each case consisted a cylindrical phantom with spheres of diameter $60~$mm, $28~$mm, and $22~$mm, and a source to background ratio of 10:1. The activity concentrations in the spheres of $200~$Bq/mL (${}^{225}$Ac) and $1~$MBq/mL (${}^{177}$Lu and ${}^{131}$I) were selected based on the expected activity concentrations in patients. The SIMIND detector and collimator parameters were identical to the point source simulations, but data were now collected at a matrix size of $128 \times 128$ pixels with $4.8~\mathrm{mm} \times 4.8~\mathrm{mm}$ resolution. 96 projection angles at 15~s/projection were collected for ${}^{177}$Lu and ${}^{131}$I, while 32 projection angles at 45000~s/projection were collected for ${}^{225}$Ac; while the ${}^{177}$Lu and ${}^{131}$I configurations were representative of a typical clinical acquisition, the ${}^{225}$Ac configuration was such that the number acquired counts was significantly higher. The energy windows used for TEW scatter correction of each isotope are shown in Table \ref{tab:e_windows}. Projection data were reconstructed with OSEM for up to 20 iterations with 8 subsets (${}^{177}$Lu and ${}^{131}$I) and up to 200 iterations with 1 subset (${}^{225}$Ac). Calibration factors were used to convert images to units of MBq/mL, and were obtained via corresponding point source simulations with the same detector configuration. The background region used to compute contrast corresponded to four 30~mm spherical ROIs placed in the warm region.

\begin{table}
\caption{\label{tab:e_windows}Energy windows employed for all image acquisitions and reconstructions in this paper to the nearest decimal in keV.}
\begin{indented}
\item[]\begin{tabular}{@{}l|lllll}
\br
Isotope&Photopeak&Lower&Upper\\
\mr
${}^{177}$Lu (MC Phantom)&187.2-228.8&169.4-187.2&228.8-252.9\\
${}^{131}$I (MC Phantom)&332.0-405.0&310.0-332.0&405.0-427.0\\
${}^{225}$Ac (MC Phantom)&396.0-484.0&374.0-396.0&484.0-506.0\\
${}^{225}$Ac (Physical Phantom)&396.0-484.0&352.0-396.0&484.0-572.0\\
${}^{225}$Ac (Patient)&396.0-484.0&358.2-395.9&484.0-506.0\\
\br
\end{tabular}
\end{indented}
\end{table}

\subsubsection{Physical Phantom Studies}

Application of the 2DG, 2D-MC, and 1DR-MC models were evaluated for reconstruction of acquired ${}^{225}$Ac physical phantom data. A cylindrical phantom with spheres of diameter $60~$mm, $28~$mm, and $22~$mm with an initial sphere activity concentration of 1.37 kBq/mL was filled at a 10:1 source to background ratio. 34 SPECT acquisitions of the phantom were taken in sequence on a Symbia T2 SPECT/CT system (Siemens Healthineers, USA) with the following settings: $128 \times 128$ pixels at $4.8~\mathrm{mm} \times 4.8~\mathrm{mm}$ resolution, 96 projection angles, high energy collimators, and 60~s acquisition time per projection; although 96 projection angles were acquired, only 32 were used since this was similar to the corresponding ${}^{225}$Ac patient acquisition in the next section. The energy windows selected for TEW additive correction are shown in Table \ref{tab:e_windows}. A blank scan with identical parameters was acquired to obtain the mean stray radiation noise in each energy window. Images were reconstructed with MLEM for up to 200 iterations. Two noise levels of the data were considered:

\begin{enumerate}
    \item One of the 34 scans was used for data reconstruction. The number of acquired counts in this scenario is approximately representative of a clinical $^{225}$Ac-PSMA-617 scan, where the patient is injected with $8~$MBq of activity \cite{ac5} and scanned for 2.5~min per projection at some time point between 0~hr and 72~hr post injection.
    \item The counts from all 34 scans were summed together and used for image reconstruction. Because the detectors had an approximately equal radial path around the phantom for each scan, this corresponds to a single scan with a high count rate.
\end{enumerate}

Spherical masks of each hot sphere were drawn using bright edges on the CT image (visible in Figure \ref{fig:real_recon}) so that the counts could be computed for each iteration number. Three additional spherical masks of diameter 50~mm were drawn in the background region in order to compute contrast.

\subsubsection{Patient Study}

A patient receiving ${}^{225}$Ac-PSMA-617 therapy with an injected activity of 8~MBq was imaged 20.5 hours post injection (first cycle). SPECT data were acquired on a Discovery 670 Pro SPECT/CT (GE Healthcare, USA) with a high energy general purpose collimator. 30 projections (15 per head) were acquired for 150~s with energy windows given by Table \ref{tab:e_windows}. Since this GE scanner has a different CDR than the Siemens scanner used in prior examples, PSF data were regenerated in SIMIND using a high energy general purpose GE collimator and another 1DR-MC model was fit as before. In GE scanners, the hexagonal bores in the collimator are rotated by $90^{\circ}$ compared to the Siemens collimator, so the angles in $\Theta_t$ were adjusted to compensate for this. The acquired data were reconstructed using 2DG, 2D-MC, and 1DR-MC PSF modeling and MLEM for up to 200 iterations. Three lesion ROIs were segmented by a physician on a pre-therapeutical PET image using the PET Edge+ tool of MIM v7.2.1 (MIM Software Inc., USA). A spherical ROI of diameter $10~$cm was placed in a low uptake region in the center of the patient so that the contrast, defined as the mean uptake ratio between the lesion ROIs and background ROI, could be obtained. The mean number of counts and contrast for each lesion ROI were then evaluated for each iteration of MLEM; in this case, before the statistics were computed for each iteration, a $3~$cm FWHM Gaussian filter was applied to the image.

\section{Results}

\begin{figure}
\includegraphics[width=\textwidth]{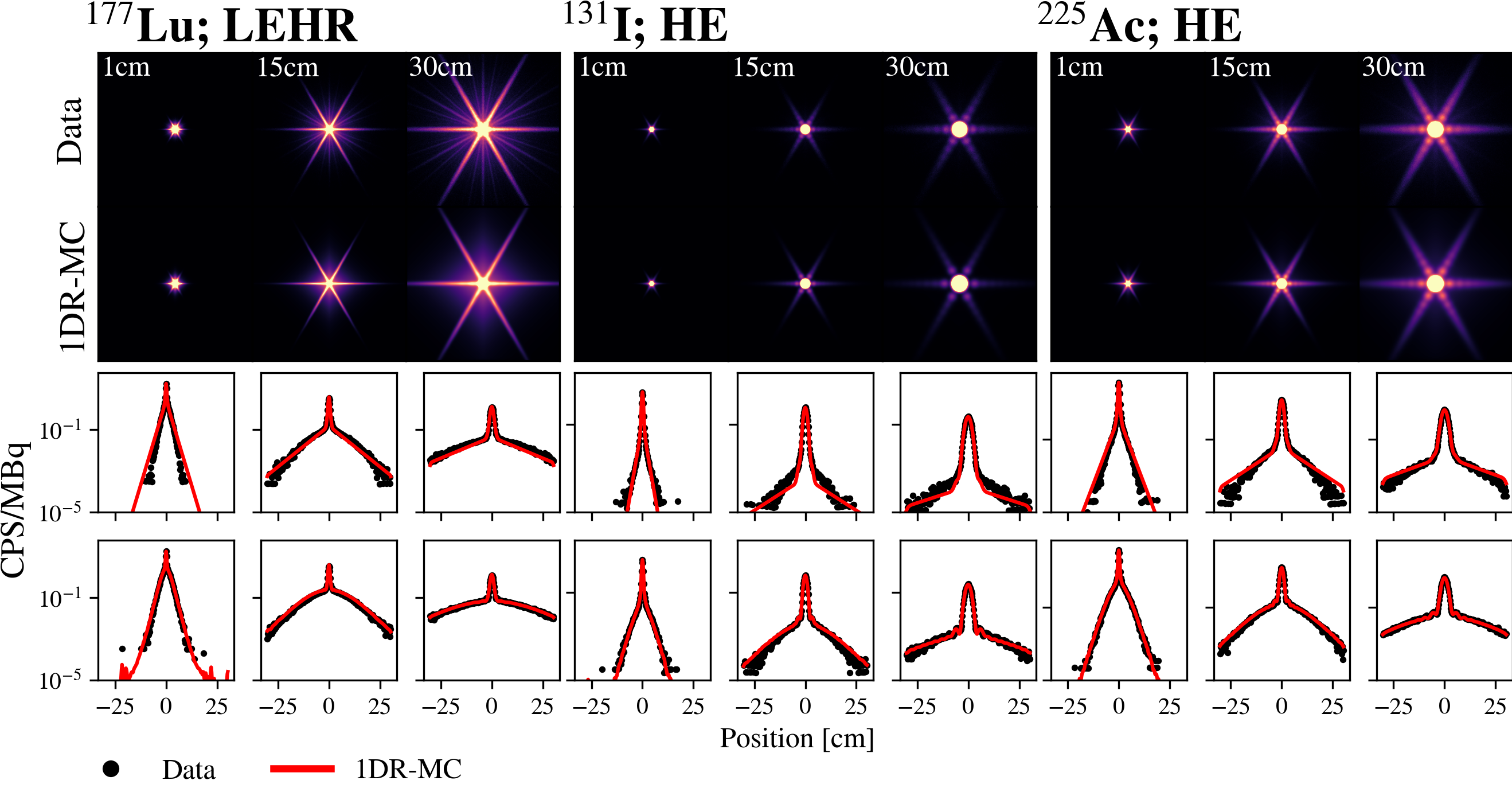}
\caption{Comparison between Monte Carlo simulated PSF data and the corresponding fit from the 1DR-MC model at different source-detector distances. From top to bottom, (i) 2D profiles of Monte Carlo PSF data, (ii) 2D profiles of 1DR-MC model, (iii) central vertical profile of Monte Carlo and fitted PSFs, (iv) central horizontal profile of Monte Carlo and fitted PSFs; the Monte Carlo data is shown in black scatter points, while the fitted data is shown as a solid red line.}\label{fig:psf_comp}
\end{figure}

Figure \ref{fig:psf_comp} compares the Monte Carlo simulated 2D-MC PSF profiles to the 1DR-MC fit obtained via Equation \ref{eq:1dr} for ${}^{177}$Lu, ${}^{131}$I, and ${}^{225}$Ac. The 1D-R model reasonable approximates all PSFs, but (i) does not always accurately capture the septal scatter component as seen in the central vertical profiles and (ii) does not capture some additional faint septal penetration tails of ${}^{177}$Lu; this is due to a limitation of the model for only including three tails at specific angles.

\begin{figure}
\centering
\includegraphics[width=\textwidth]{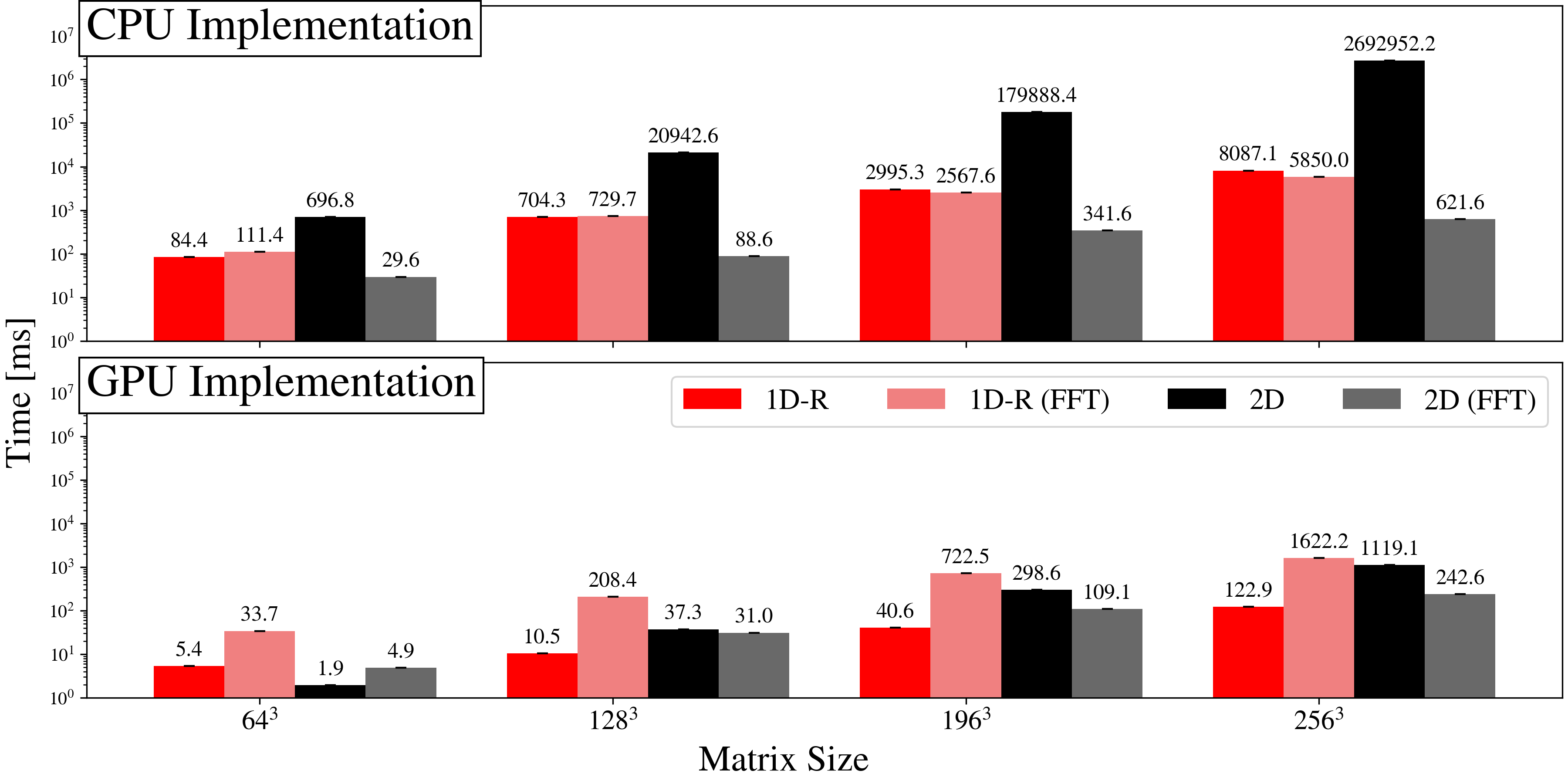}
\caption{CPU and GPU required run time of the 1DR-MC and 2D PSF models for different matrix sizes. Each implementation is evaluated using standard and FFT based convolution. The time shown on top of each bar in units of ms.}\label{fig:timing}
\end{figure}

Timing benchmarks for PSF modeling using the 1DR-MC, 1DR-MC (FFT), 2D-MC, and 2D-MC (FFT) are shown in Figure \ref{fig:timing}. CPU implementation yields no benefits with the proposed 1DR-MC model, and performs fastest with FFT-based 2D convolutions and slowest using regular 2D convolutions. The GPU implementation is faster than the CPU implementation for all methods, and yields computational benefits when the proposed 1DR-MC model is used. With a matrix size of $128^3$, use of the 1DR-MC method is over three times faster than 2D-MC (FFT) PSF modeling. For small matrix sizes ($64^3$) there is no computational speed-up using the 1DR-MC method, and for large matrix sizes ($256^3$) the relative time difference between the 1DR-MC and 2D-MC (FFT) begins to decrease, approaching only a two times speed advantage.

\begin{figure}
\centering
\includegraphics[width=\textwidth]{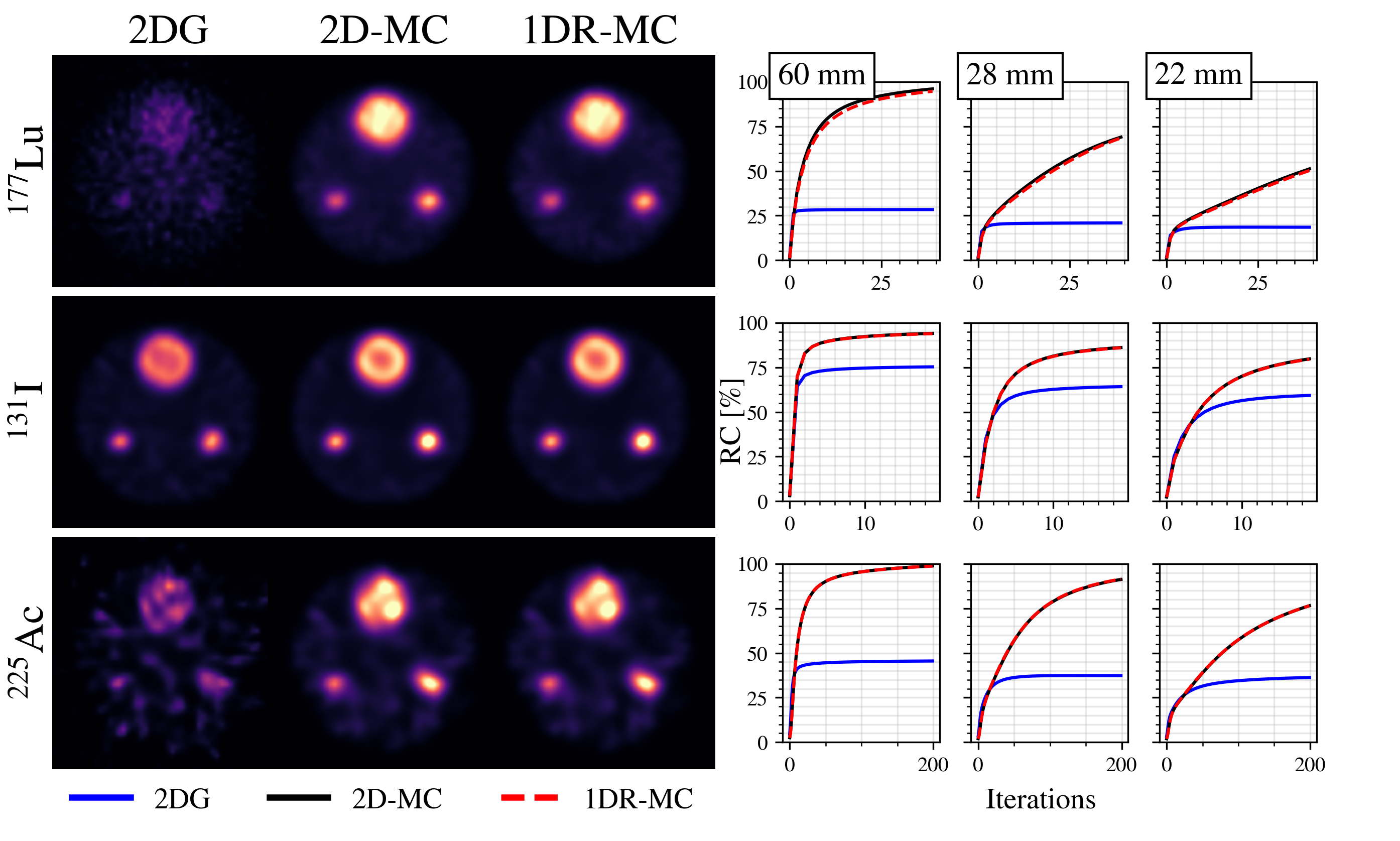}
\caption{Reconstructions of the MC simulations (left) and corresponding RCs for each of the three spheres at each iteration (right). The displayed images correspond to reconstructions after 40 iterations (${}^{177}$Lu), 10 iterations (${}^{131}$I) and 100 iterations (${}^{225}$Ac)}\label{fig:MC_recons}
\end{figure}

Reconstructed MC simulations of the ${}^{177}$Lu, ${}^{131}$I, and ${}^{225}$Ac spherical phantoms are shown in Figure \ref{fig:MC_recons}, while corresponding RC and contrast statistics (for the last iteration of reconstruction) are shown in Table \ref{tab:rc_and_contrast_phan}. The 1DR-MC and 2D-MC models yield similar RCs in all cases, and are larger than those obtained using the 2DG model. 

\begin{table}
\caption{\label{tab:rc_and_contrast_phan}Recovery coefficients (or counts) and contrast for various regions of interest in all reconstruction examples using (i) traditional 2D Gaussian PSF modeling (2DG), (ii) 2D kernels (2D-MC), and (iii) the proposed 1DR-MC model that aims to to be consistent with the 2D-MC model while reducing reconstruction times. The low count ${}^{225}$Ac physical phantom corresponds to the first of the 34 acquisitions.}
\begin{indented}
\item[]\begin{tabular}{@{}l|lll|lll}
\br
\centre{7}{${}^{177}$Lu MC Phantom}\\
\br
&\centre{3}{Contrast}&\centre{3}{RC [\%]}\\
Model&60~mm&28~mm&22~mm&60~mm&28~mm&22~mm\\
\mr
2DG & 2.68 & 2.00 & 1.73 & 28.4 & 21.2 & 18.3 \\
2D-MC & 7.85 & 5.72 & 4.23 & 95.9 & 69.9 & 51.7 \\
1DR-MC & 8.19 & 6.01 & 4.40 & 94.6 & 69.4 & 50.8 \\
\br
\centre{7}{${}^{131}$I MC Phantom}\\
\br
&\centre{3}{Contrast}&\centre{3}{RC [\%]}\\
Model&60~mm&28~mm&22~mm&60~mm&28~mm&22~mm\\
\mr
2DG & 8.08 & 6.94 & 6.22 & 75.4 & 64.7 & 58.0 \\
2D-MC & 10.41 & 9.60 & 8.74 & 94.1 & 86.7 & 79.0 \\
1DR-MC & 10.31 & 9.49 & 8.64 & 94.0 & 86.5 & 78.8 \\
\br
\centre{7}{${}^{225}$Ac MC Phantom}\\
\br
&\centre{3}{Contrast}&\centre{3}{RC [\%]}\\
Model&60~mm&28~mm&22~mm&60~mm&28~mm&22~mm\\
\mr
2DG & 4.55  & 3.73  & 3.63  & 45.5 & 37.3 & 36.2 \\
2D-MC & 9.11  & 8.42  & 7.05  & 98.8 & 91.2 & 76.5 \\
1DR-MC & 9.03  & 8.36  & 6.99  & 98.7 & 91.4 & 76.5 \\
\br
\centre{7}{${}^{225}$Ac Physical Phantom (Low Count)}\\
\br
&\centre{3}{Contrast}&\centre{3}{Counts}\\
Model&60~mm&28~mm&22~mm&60~mm&28~mm&22~mm\\
\mr
2DG  & 4.01 & 3.68 & 3.32 & 0.0342  & 0.0314  & 0.0283 \\
2D-MC & 4.85 & 4.80 & 4.36 & 0.0640  & 0.0634  & 0.0575 \\
1DR-MC & 4.88 & 4.80 & 4.36 & 0.0643  & 0.0632  & 0.0574\\
\br
\centre{7}{${}^{225}$Ac Physical Phantom (High Count)}\\
\br
&\centre{3}{Contrast}&\centre{3}{Counts}\\
Model&60~mm&28~mm&22~mm&60~mm&28~mm&22~mm\\
\mr
2DG & 4.68  & 4.31  & 3.70 & 1.24  & 1.14  & 0.98 \\
2D-MC & 8.15  & 7.83  & 5.98 & 2.52  & 2.42  & 1.85 \\
1DR-MC & 8.05  & 7.69  & 5.92 & 2.56  & 2.45  & 1.88 \\
\br
\centre{7}{${}^{225}$Ac Patient}\\
\br
&\centre{3}{Contrast}&\centre{3}{Counts}\\
Model&Les 1&Les 2& Les 3& Les 1& Les 2 &Les 3\\
\mr
2DG & 2.32  & 2.28  & 2.80  & 0.0405 & 0.0267 & 0.0660 \\
2D-MC & 3.72  & 3.50  & 4.86  & 0.0405 & 0.0273 & 0.0653 \\
1DR-MC & 3.74  & 3.50  & 4.92  & 0.0246 & 0.0254 & 0.0355 \\
\br
\end{tabular}
\end{indented}
\end{table}

\begin{figure}
\centering
\includegraphics[width=\textwidth]{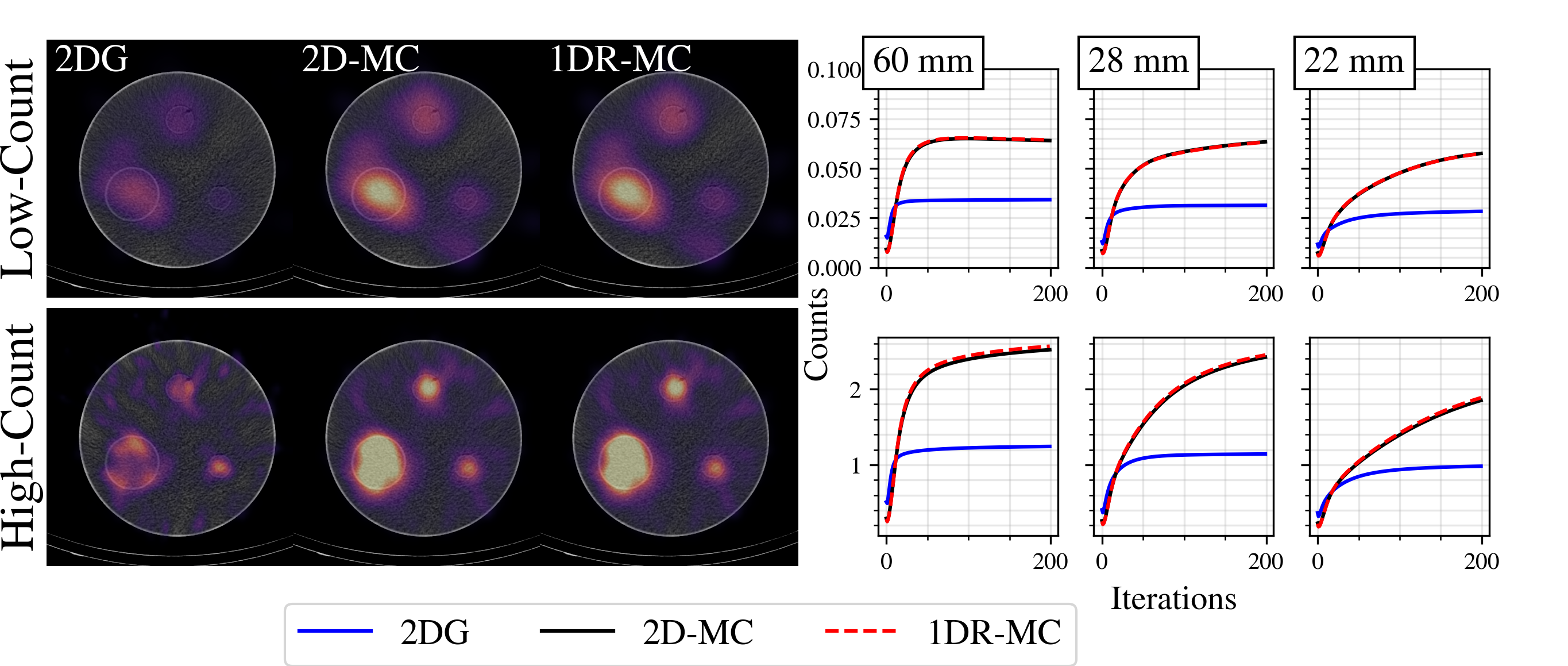}
\caption{Reconstruction of physical ${}^{225}$Ac phantom using various PSF models. Reconstructed SPECT and CT images are superimposed; the boundaries of each sphere can be seen on the CT. The displayed images correspond to reconstruction after 100 iterations; the low count images were additionally post-smoothed using a 3D Gaussian function with a 3~cm FWHM. The plots on the right show the average counts per voxel in the three spheres.}\label{fig:real_recon}
\end{figure}

Reconstructions of physical ${}^{225}$Ac phantom images are shown in Figure \ref{fig:real_recon}, and quantitative statistics for (i) the first low count acquisition and (ii) the high count acquisition after 200 iterations are shown in Table \ref{tab:rc_and_contrast_phan}. Images generated from the 1DR-MC and 2D-MC methods are almost indistinguishable qualitatively. The quality metrics in Table \ref{tab:rc_and_contrast_phan} demonstrate similar counts and contrast for the 2D-MC and 1DR-MC models, and significantly lower counts and contrast for the 2DG model. The time required for 200 iterations of MLEM for the low count data was $77~$s (2DG PSF), $299~$s (1DR-MC PSF) and $748~$s (2D-MC PSF). All 34 low count acquisitions were also reconstructed and the 2DG, 2D-MC, and 1DR-MC methods were quantitatively compared; the differences in total counts between the 2D-MC and 1DR-MC were $(-0.01 \pm 1.20)\%$, $(0.67 \pm 0.23)\%$, and $(-0.75\pm 1.25)\%$ for smallest to largest sphere respectively. As a demonstration of how minor these differences are, the variability of counts between the 34 acquisitions for the 2D-MC model were $61.3\%$, $38.5\%$, and $11.1\%$ from smallest to largest sphere, respectively.

\begin{figure}
\centering
\includegraphics[width=\textwidth]{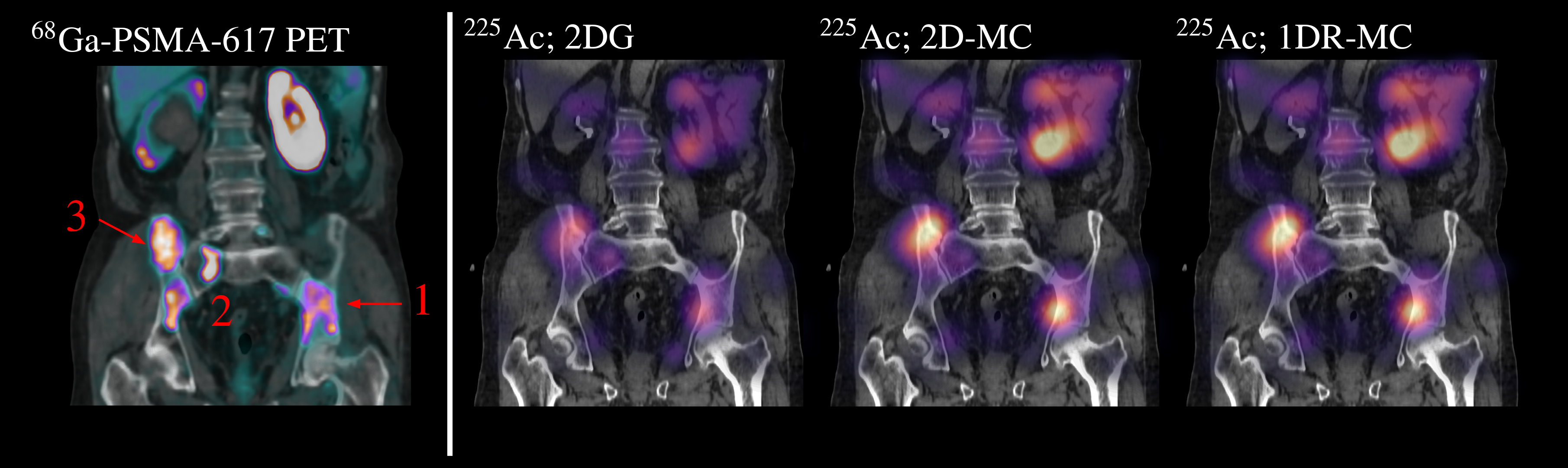}
\includegraphics[width=\textwidth]{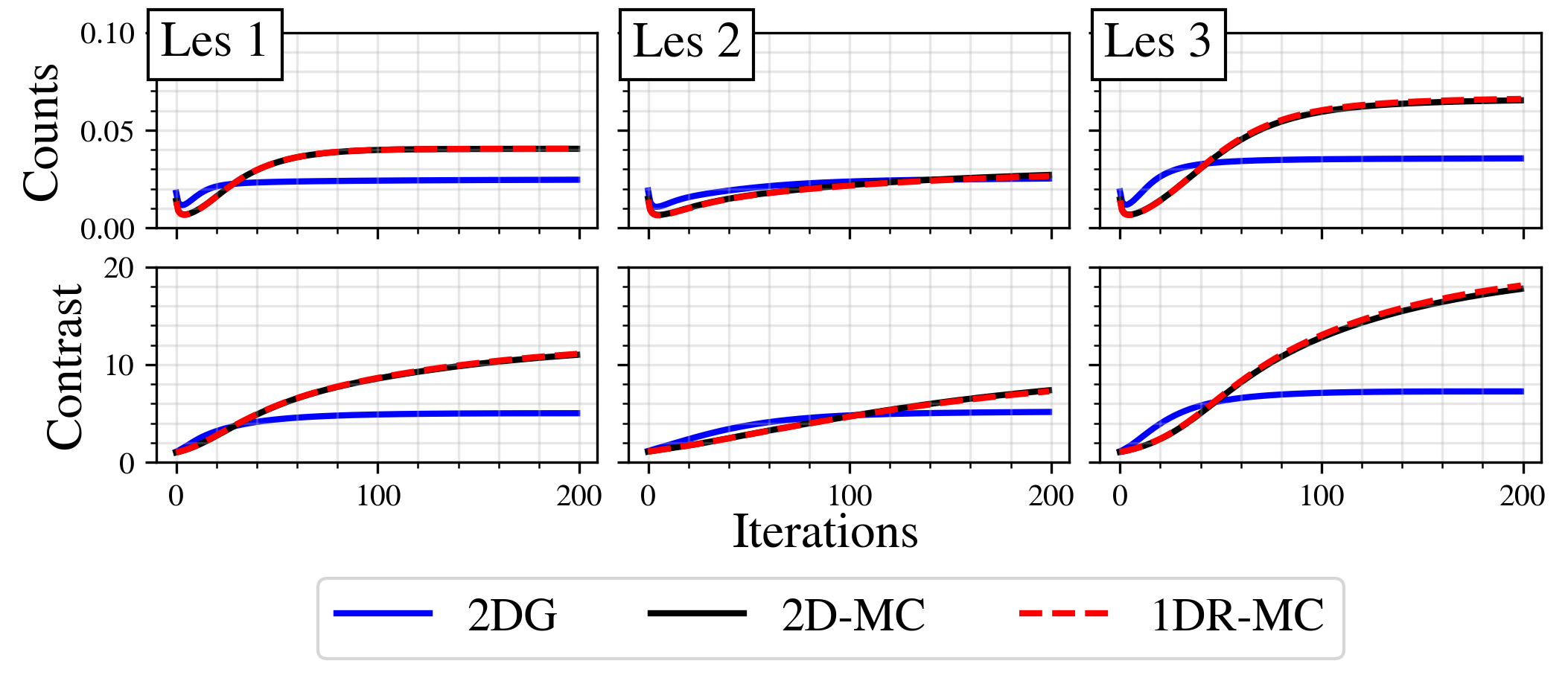}
\caption{Coronal slice of a ${}^{68}$Ga-PSMA-617 PET image (top row, left) and associated ${}^{225}$Ac-PSMA-617 SPECT data (top row, right), counts in bone lesion ROIs as a function of iteration number (middle row) and signal to background ratios in the bone lesion ROIs (bottom row). ${}^{225}$Ac images were reconstructed in PyTomography using the 2DG, 2D-MC, and 1DR-MC PSF models. For computation of the data in the line plots, the image at each iteration was post-filtered using a 3~cm FWHM 3D Gaussian filter before the statistic was computed. Arrows pointing to the three bone lesions are shown in red (lesion 2 is directly on top of its annotation). There was no uptake in the right kidney due to organ failure.}\label{fig:patient_recon}
\end{figure}

PyTomography reconstructed ${}^{225}$Ac-PSMA-617 patient images are shown in Figure \ref{fig:patient_recon}, along with a reference ${}^{68}$Ga-PSMA-617 PET scan acquired on a GE Discovery IQ scanner and reconstructed using Q.Clear ($\beta=450$). The time required for 200 iterations of MLEM was $75.1$~s (2DG), $272.0~$s (1DR-MC) and $704.0~$s (2D-MC). Qualitatively, images reconstructed using the Gaussian PSF can be observed to have less counts in the lesion ROIs. Quantitative results are shown as line plots in Figure \ref{fig:patient_recon}, and quality metrics for iteration 200 are shown in Table \ref{tab:rc_and_contrast_phan}. These results demonstrate (i) indistinguishable images and nearly identical image metrics between the 1DR-MC and 2D-MC models and (ii) higher contrast and count uptake for 2D-MC and 1DR-MC models relative to the 2DG model for all three lesions.

\section{Discussion}

This work proposes a 1DR-MC PSF model for high-energy SPECT reconstruction that demonstrates consistency with 2D-MC models, while also increasing computational efficiency for GPU-based reconstruction of ${}^{177}$Lu, ${}^{131}$I, and ${}^{225}$Ac at typical SPECT matrix sizes. The model was created using the publicly shared SPECTPSFToolbox component of the PyTomography project; it is hoped that the shared toolbox and corresponding tutorials will permit the nuclear medicine community to develop custom PSF models for new isotope/collimator configurations as research in novel RPTs and imaging techniques continues to grow.

As shown in Figure \ref{fig:timing}, the proposed 1DR-MC only yields computational benefits when image reconstruction is performed on GPU. If reconstruction is performed on CPU, then PSF modeling is faster when implemented with FFT-based 2D convolutions. As discussed in Section 2.1.2, the kernel sizes were fixed to $N-1 \times N-1$ for all $d$. For small $d$, where the PSF is also smaller, a kernel of this size might be excessive and could unnecessarily increase the computational time required. Future research exploring the reduction of computational time might experiment with using a source-detector distance dependent kernel size $N(d)$ for that matches the size of the PSF; this would further reduce the computational time required. 

While the proposed 1DR-MC model is not able to perfectly capture all the PSF features in Figure \ref{fig:psf_comp}, it yields qualitatively similar reconstructions of the MC data in Figure \ref{fig:MC_recons}. Furthermore, as shown in Table \ref{tab:rc_and_contrast_phan}, the RCs and contrast of the 1DR-MC and 2D-MC models are similar and always significantly greater than the 2DG model. This demonstrates the superior quantitative performance of the MC-based models, and justification for using the 1DR-MC over the 2D-MC method due to faster reconstruction times.

For the physical ${}^{225}$Ac phantom and patient use cases shown in Figures \ref{fig:real_recon} and \ref{fig:patient_recon}, the 2D-MC and 1DR-MC models yield similar count quantification, especially when compared to the 2DG model. The quality metrics for 200 iterative updates shown in Table \ref{tab:rc_and_contrast_phan} further highlight this consistency. In the physical phantom high count case, use of the 1DR-MC method compared to the 2DG method increased the mean counts the spheres by factors of 2.06, 2.14, and 1.92 for the largest to smallest sphere respectively; this is somewhat similar to the MC study where the RCs increased by 2.17, 2.45, and 2.11 for spheres of the same size. While calibration factors were not available for the physical data, inference from the MC results suggest that the RCs of the physical data were significantly improved.

In the low count scenario of the physical phantom study, which approximately represents a clinical count rate, the percent difference in the three spheres between the 2D-MC and 1DR-MC models were small compared to the variability between the thirty-four separate acquisitions. This highlights that any systematic errors between use of the 2D-MC and 1DR-MC model are negligible compared to the precision of the imaging protocol, and thus that the 1DR-MC model may be suitable for standard ${}^{225}$Ac image reconstruction. In the patient example, it was similarly found that (i) the differences between the 2D-MC and 1DR-MC models in the bone lesions were small and (ii) the 2DG model had significantly reduced contrast relative to the 2D-MC and 1DR-MC models. 

The functional form of Equations \ref{eq:gauss_comp}, \ref{eq:tail_comp}, and \ref{eq:bkg_comp} were selected because they produced a reasonable approximation of the PSF, but they can be further developed. Use of rotation operations with the component $K_B^{(x;d;b)}K_B^{(y;d;b)}$ may permit more radial symmetry in the SSRF. Substitution of $K_G^{(x;d;b)}K_G^{(y;d;b)}$ with a small 2D kernel that represents the true aperture function of the GRF may also improve the accuracy of the PSF model. The ${}^{177}$Lu PSF data in Figure \ref{fig:psf_comp} reveals additional dim tails; these could be modeled by including another independent tail component in the model. A trade-off for adding additional features to the model, however, is an increase in computation time. The SPECTPSFToolbox Python library released along with this paper contains tutorials demonstrating how to create custom and fit parameterized PSF models. Users can obtain point source data using Monte Carlo programs (such as GATE \cite{gate} and SIMIND) or by using real scanners, and fit corresponding models to the acquired data. The models can then be imported to PyTomography for customized and scanner specific SPECT reconstruction. Users can independently evaluate the trade-off between adding model features, impact on reconstructed images, and increase in computation time.

Since GPU-accelerated reconstruction using the proposed 1DR-MC model reduces reconstruction times by more than a factor of two compared to the 2D-MC model, it may be preferable for reconstruction of ${}^{225}$Ac data for typical SPECT matrix sizes. Both the 1DR-MC and 2D-MC models, however, are only applicable when Equation \ref{eq:model_approx} is used for SPECT system matrix modeling; this equation relies on the assumption that the attenuation probabilities vary little across the PSF. Due to the large spatial extent of the PSFs shown here, this assumption may be invalid, and Equation \ref{eq:model_full} may instead be required for accurate reconstruction. This assumption largely depends on the density of the object being scanned. If the object had a region of high density which only rays on the outer edges of the PSF passed through, then the equal attenuation path assumption would be invalid. For patients and phantoms, where the density is roughly constant throughout the field of view, application of Equation \ref{eq:model_approx} may be permitted. Meanwhile, the study of different density configurations and the applicability of Equation \ref{eq:model_approx} remains to be investigated in future work.

\section{Conclusion}

In conclusion, a computationally efficient implementation of high energy CDR modeling in SPECT imaging was developed and tested on ${}^{177}$Lu, ${}^{131}$I, and $^{225}$Ac reconstructions. The technique was implemented using the open source reconstruction library PyTomography, and was shown to speed up reconstruction times by more than a factor of two compared to conventional, 2D convolution based methods ($272.0$ s vs.\ $704.0$ s for patient data). Furthermore, the proposed 1DR-MC method yielded near identical results to the reference 2D-MC approach, and the small differences for ${}^{225}$Ac were insignificant compared to the differences between reconstructions of separate noise realizations. The SPECTPSFToolbox python library was developed and publicly shared to permit others in the nuclear medicine community to develop custom isotope/collimator PSF models for use in the open-source reconstruction library PyTomography.

\section{Acknowledgements}

This work was supported by the Natural Sciences and Engineering Research Council of Canada (NSERC) CGS D Award 569711 and Discovery Grant RGPIN-2019-06467, as well as computational resources and services provided by Microsoft AI for Health.

\section{References}


\end{document}